\documentclass[12pt]{article}
\usepackage{graphicx}
\def\hybrid{\topmargin 0pt      \oddsidemargin 0pt
        \headheight 0pt \headsep 0pt
        \voffset=-0.5cm
        \textwidth 6.25in       
        \textheight 9.5in       
        \marginparwidth 0.0in
        \parskip 5pt plus 1pt   \jot = 1.5ex}
\catcode`\@=11
\def\marginnote#1{}

\newcount\hour
\newcount\minute
\newtoks\amorpm
\hour=\time\divide\hour by60
\minute=\time{\multiply\hour by60 \global\advance\minute by-\hour}
\edef\standardtime{{\ifnum\hour<12 \global\amorpm={am}%
        \else\global\amorpm={pm}\advance\hour by-12 \fi
        \ifnum\hour=0 \hour=12 \fi
        \number\hour:\ifnum\minute<10 0\fi\number\minute\the\amorpm}}
\edef\militarytime{\number\hour:\ifnum\minute<10 0\fi\number\minute}

\def\draftlabel#1{{\@bsphack\if@filesw {\let\thepage\relax
   \xdef\@gtempa{\write\@auxout{\string
      \newlabel{#1}{{\@currentlabel}{\thepage}}}}}\@gtempa
   \if@nobreak \ifvmode\nobreak\fi\fi\fi\@esphack}
        \gdef\@eqnlabel{#1}}
\def\@eqnlabel{}
\def\@vacuum{}
\def\draftmarginnote#1{\marginpar{\raggedright\scriptsize\tt#1}}
\def\draftlabel#1{{\@bsphack\if@filesw {\let\thepage\relax
   \xdef\@gtempa{\write\@auxout{\string
      \newlabel{#1}{{\@currentlabel}{\thepage}}}}}\@gtempa
   \if@nobreak \ifvmode\nobreak\fi\fi\fi\@esphack}
        \gdef\@eqnlabel{#1}}
\def\@eqnlabel{}
\def\@vacuum{}
\def\draftmarginnote#1{\marginpar{\raggedright\scriptsize\tt#1}}

\def\draft{\oddsidemargin -.5truein
        \def\@oddfoot{\sl preliminary draft \hfil
        \rm\thepage\hfil\sl\today\quad\militarytime}
        \let\@evenfoot\@oddfoot \overfullrule 3pt
        \let\label=\draftlabel
        \let\marginnote=\draftmarginnote
   \def\@eqnnum{(\theequation)\rlap{\kern\marginparsep\tt\@eqnlabel}%
\global\let\@eqnlabel\@vacuum}  }

\def\numberbysection{\@addtoreset{equation}{section}
        \def\theequation{\thesection.\arabic{equation}}}

\def\underline#1{\relax\ifmmode\@@underline#1\else
        $\@@underline{\hbox{#1}}$\relax\fi}

\def\titlepage{\@restonecolfalse\if@twocolumn\@restonecoltrue\onecolumn
     \else \newpage \fi \thispagestyle{empty}\c@page\z@
        \def\thefootnote{\fnsymbol{footnote}} }

\def\endtitlepage{\if@restonecol\twocolumn \else  \fi
        \def\thefootnote{\arabic{footnote}}
        \setcounter{footnote}{0}}  
\relax

\numberbysection
\hybrid

\newfont{\Bbb}{msbm10 scaled 1\@ptsize00}
\newfont{\Bbbb}{msbm7 scaled 1\@ptsize00}
\newcommand{\CC}{\mbox{\Bbb C}}

\newcommand{\DDD}{\raise-1pt\hbox{$\mbox{\Bbbb D}$}}

\newcommand{\II}{\mbox{\Bbb I}}

\newcommand{\UUU}{\raise-1pt\hbox{$\mbox{\Bbbb U}$}}

\newcommand{\ZZ}{\mbox{\Bbb Z}}
\newcommand{\z}{\raise-1pt\hbox{$\mbox{\Bbbb Z}$}}

\def\beq{\begin{equation}}
\def\eeq{\end{equation}}
\def\p{\partial}

\begin{document}

\begin{titlepage}

\title{The master $T$-operator for vertex
models with trigonometric $R$-matrices as
classical tau-function}

\author{A.~Zabrodin
\thanks{Institute of Biochemical Physics,
4 Kosygina st., 119334, Moscow, Russia and ITEP, 25
B.Cheremushkinskaya, 117218, Moscow, Russia and
NRU-HSE, Vavilova str. 7, 117312 Moscow, Russia}}

\date{May 2012}
\maketitle

\vspace{-7cm} \centerline{ \hfill ITEP-TH-17/12} \vspace{7cm}

\begin{abstract}

The construction of the master $T$-operator recently
suggested in \cite{AKLTZ11} is applied to integrable vertex models
and associated quantum spin chains with trigonometric
$R$-matrices. The master $T$-operator is a generating function
for commuting transfer matrices of integrable vertex models
depending on infinitely many parameters.
At the same time it turns out to be the tau-function 
of an integrable hierarchy
of classical soliton equations
in the sense that it satisfies the
the same bilinear Hirota equations.
The class of solutions of the Hirota equations
that correspond to eigenvalues of the master $T$-operator
is characterized
and its relation to the classical Ruijsenaars-Schneider
system of particles is discussed.

\end{abstract}

\end{titlepage}

\section{Introduction}

The master $T$-operator was recently
introduced in \cite{AKLTZ11}. It is a generating function
for commuting transfer matrices of integrable vertex models
and associated quantum spin chains which unifies the transfer matrices
on all levels of the nested Bethe ansatz and Baxter's
$Q$-operators in one commuting family.
It was also proven in \cite{AKLTZ11} that the master $T$-operator,
as a function of infinitely many auxiliary parameters
(one of which being the usual spectral parameter),
satisfies the same hierarchy of bilinear Hirota equations
as the classical $\tau$-function does. Since the operator-valued generating functions
commute for all values of the auxiliary parameters,
there is no problem with their ordering in the bilinear equations.

A similarity between quantum transfer matrices and classical
$\tau$-functions was first pointed out in \cite{KLWZ97}
(see also \cite{Z97}),
where a discrete integrable dynamics in the space of commuting
integrals of motion of a quantum integrable model was introduced.
This classical dynamics was identified with
the discrete 3-term Hirota equation with special
boundary conditions. The diagonalization of transfer matrices
by means of the nested Bethe
ansatz technique was shown to be
equivalent to an ``undressing'' chain of B\"acklund
transformations for the discrete Hirota equation.
Later this approach was extended to
supersymmetric integrable models \cite{KSZ08}.
An essential further step was
made in the important paper \cite{KLT10},
where an operator realization of the B\"acklund flow describing
the ``undressing'' process was constructed for
generalized quantum spin chains with rational
$GL(N)$-invariant $R$-matrices. In fact the master $T$-operator
was already used implicitly in that construction.
A more explicit and more general definition was given
in \cite{AKLTZ11}.

In this paper we review the construction of \cite{AKLTZ11}
trying to avoid technical details. Here we deal with the
class of integrable lattice vertex models of statistical mechanics
with trigonometric $R$-matrices.
The main claim is that the master $T$-operator for these models
is a $\tau$-function of the
classical MKP hierarchy.

We also
characterize the class of solutions of the Hirota equations
that correspond to eigenvalues of the master $T$-operator
and make explicit the close connection with the classical
Ruijsenaars-Schneider
system of particles \cite{RuijSch}
which emerges as the dynamical system for
zeros of the (eigenvalues of) the master $T$-operator.
In an equivalent way, the connection emerges from the
Baker-Akhiezer function for the Ruijsenaars-Schneider
system which generates the algebra of commuting operators
(transfer matrices) for the vertex model (the Bethe
algebra).
As is well known,
the Ruijsenaars-Schneider model 
contains
the Calogero-Moser system of particles
as a limiting case. In this connection
let us note that a similar relation between the quantum Gaudin
model (which can be regarded as a degeneration
of quantum spin chains or vertex models
with rational $R$-matrices) and classical Calogero-Moser
system was found in \cite{MTV} from a different
reasoning.

\section{The transfer matrices}

We consider generalized quantum integrable vertex models
with trigonometric $R$-matrix. The simplest $R$-matrix is the
operator in $\CC ^N \otimes \CC ^N$ of the form
\beq\label{R1}
\begin{array}{c}
\displaystyle{
{\sf R}(u)=(e^{\gamma (u+1)}\! -\! e^{-\gamma (u+1)})
\sum_{a=1}^{N}e_{aa}\otimes e_{bb}+
(e^{\gamma u}-e^{-\gamma u})\sum_{1\leq a\neq b\leq N}
e_{aa}\otimes e_{bb}}
\\ \\
\displaystyle{+\, (e^{\gamma}-e^{-\gamma})
\sum_{1\leq a\neq b\leq N}e^{\mbox{\scriptsize sign}(b-a)\gamma u}
e_{ab}\otimes e_{ba}}.
\end{array}
\eeq
Here $u\in \CC$ is the spectral parameter and
$e_{ab}$ denotes the $N\times N$ matrix with $1$
in position $(a,b)$ and $0$ elesewhere.
The deformation (anisotropy) parameter $\gamma$ is
assumed to be such that $q=e^{\gamma}$
is not a root of unity.
Following the tradition, we call this $R$-matrix
{\it trigonometric}
although the coefficients are
hyperbolic functions of $u$ like $\sinh \gamma u$.
Let $V_i=\CC ^N$ be copies of the space $\CC^N$, then
by ${\sf R}_{ij}(u)$ denote
the $R$-matrix acting in $V_i \otimes V_j$.
The $R$-matrix (\ref{R1}) satisfies the Yang-Baxter
equation
\beq\label{R2}
{\sf R}_{12}(u_1 -u_2){\sf R}_{13}(u_1 -u_3)
{\sf R}_{23}(u_2 -u_3)=
{\sf R}_{23}(u_2 -u_3){\sf R}_{13}(u_1 -u_3)
{\sf R}_{12}(u_1 -u_2),
\eeq
where the both sides are operators in $V_1\otimes V_2 \otimes V_3$.
For any diagonal $N\times N$ matrix $g$ set $g_1 =g\otimes \II$,
$g_2 = \II \otimes g$, then the $R$-matrix commutes with
$g_1 g_2$:
\beq\label{g1}
{\sf R}_{12}(u)\, g_1 g_2=g_1 g_2 \, {\sf R}_{12}(u).
\eeq
This property will be referred to as $g$-invariance of the $R$-matrix.

Fix a diagonal matrix $g=\mbox{diag}\, (p_1, p_2, \ldots , p_N)$.
We call it the twist matrix.
Below we assume that all $p_i \in \CC$ are in general position,
i.e., $p_i/p_j \neq e^{2\gamma n}$ for any $i\neq j$ and any integer $n$.
The transfer matrix for the
vertex model with twisted boundary conditions
and with inhomogeneity parameters $u_i$ at each site
is defined as
\beq\label{R3}
T(u)=\mbox{tr}_0 \Bigl ( {\sf R}_{10}(u-u_1){\sf R}_{20}(u-u_2)\ldots
{\sf R}_{L0}(u-u_L)\, g\Bigr ).
\eeq
The $R$-matrices and $g$
are mulitiplied as matrices in the common space $V_0$
(the auxiliary space). Trace $\mbox{tr}_0$
is taken in the auxiliary space.
The result is an operator acting
in the tensor product of vector representations
${\cal H}=\otimes_{j=1}^{L}V_j =
(\CC^N)^{\otimes L}$ (the quantum space).
Formally, our setting includes also models with higher representations
at the sites because they can be obtained by ``fusing'' several vector
representations with properly chosen parameters $u_i$.
By construction, the operator (\ref{R3})
is a Laurent polynomial in $e^{\gamma u}$.

It follows from the Yang-Baxter equation
and from the $g$-invariance of the $R$-matrix
that the transfer matrices
for models with the same $\gamma$
and $g$ commute for all $u$
and can be diagonalized simultaneously.
Their diagonalization is the basic problem of the theory of
vertex models. The standard method is the nested Bethe
ansatz technique.

The full commutative family of operators
in the quantum space is in general larger than the one generated by
coefficients of
$T(u)$. The algebraic construction of higher commuting transfer
matrices essentially relies on representation theory of
the $q$-deformed algebras $U_q(\widehat{gl}(N))$ and
$U_q({gl}(N))$ (see, e.g., \cite{CP,Rosso,KS,ACDFR06}).

The algebra $U_q({gl}(N))$ has generators $L^{+}_{ab}$
with $1\leq a\leq b\leq N$ and $L^{-}_{ab}$
with $1\leq b\leq a\leq N$ such that $L^{+}_{aa}L^{-}_{aa}=
L^{-}_{aa}L^{+}_{aa}=1$. Combining them into matrices
$$\displaystyle{{\sf L}^{+}= \sum_{a\leq b} e_{ab}\otimes L^{+}_{ab}},
\quad
\displaystyle{{\sf L}^{-}= \sum_{a\geq b} e_{ab}\otimes L^{-}_{ab}}
$$
with $U_q({gl}(N))$-valued matrix elements,
one can represent the defining relations
of the algebra in the form \cite{FRT}
$$
{\sf R}_{12}{\sf L}^{\pm}_{1}{\sf L}^{\pm}_{2}=
{\sf L}^{\pm}_{2}{\sf L}^{\pm}_{1}{\sf R}_{12},
\quad
{\sf R}_{12}{\sf L}^{+}_{1}{\sf L}^{-}_{2}=
{\sf L}^{-}_{2}{\sf L}^{+}_{1}{\sf R}_{12}
$$
with the $u$-independent $R$-matrix
$$
{\sf R}_{12}=
\lim_{e^{\gamma u}\to \infty}(e^{-\gamma u}{\sf R}(u))=
q\sum_{a=1}^{N}e_{aa}\otimes e_{bb}+\!\!
\sum_{1\leq a\neq b\leq N}\!\!
e_{aa}\otimes e_{bb}+(q\! -\! q^{-1})\!\!\!
\sum_{1\leq a\leq b\leq N} \!\! e_{ab}\otimes e_{ba}.
$$
The diagonal elements $L_{aa}^{\pm}$ can be understood as exponents
of the commuting Cartan generators ${\sf h}_a$:
$$
L_{aa}^{\pm}=q^{\pm {\sf h}_a}.
$$

Let $\pi_{\lambda}$ be the irreducible finite-dimensional
representation of $U_q({gl}(N))$ with the highest weight
$\lambda =(\lambda_1 ,\lambda_2 , \ldots ,
\lambda_N)$ such that $\lambda_i \in \ZZ_+$,
$\lambda_1 \geq \lambda_2 \geq \ldots \geq
\lambda_N$. The highest weight vector ${\sf v}$ obeys
$$
L^{-}_{ab}{\sf v}=0, \quad a>b, \quad \quad
L^{-}_{aa}{\sf v}=q^{-\lambda_a}{\sf v}.
$$
The representation space $V^{(\lambda )}$
is generated by repeated action of
the generators $L^{+}_{ab}$ on the highest weight vector
and subsequent factorizing (see \cite{Rosso} for details).
These representations are
$q$-deformations of the highest weight finite-dimensional
representations of $U(gl(N))$. The highest weights are
naturally identified with Young diagrams (partitions)
$\lambda$.

The representation $\pi_{(1)}$ corresponding to
the one-box diagram is the vector representation
in $\CC ^N$. On the Cartan generators ${\sf h}_a$
introduced above it looks exactly like for the
usual non-deformed algebra $gl(N)$:
$
\pi_{(1)}({\sf h}_a) = e_{aa}.
$
Given the
diagonal matrix $g=\mbox{diag}\, (p_1, p_2, \ldots , p_N)$, we set
\beq\label{twistel}
{\sf g}=p_1^{{\sf h}_1}p_2^{{\sf h}_2}\ldots
p_N^{{\sf h}_N},
\eeq
then $g=\pi_{(1)}({\sf g})$.

The $R$-matrix ${\sf R}^{\lambda}(u)$
acting in $\CC ^N \otimes V^{(\lambda )}$ is
\beq\label{R4}
{\sf R}^{\lambda}(u)=e^{\gamma u}
\sum_{a\leq b}e_{ab}\otimes \pi_{\lambda}(L_{ab}^{+})-
e^{-\gamma u}
\sum_{a\geq b}e_{ab}\otimes \pi_{\lambda}(L_{ab}^{-}).
\eeq
The $R$-matrices ${\sf R}^{\lambda}(u)$,
${\sf R}^{\mu}(u)$ are intertwined by a more general
$R$-matrix $R^{\lambda \mu}(u)$ which acts in
$V^{(\lambda )}\otimes V^{(\mu )}$:
\beq\label{R5}
{\sf R}_{12}^{\lambda \mu}(u_1 -u_2){\sf R}^{\lambda}_{13}(u_1-u_3)
{\sf R}^{\mu}_{23}(u_2-u_3)={\sf R}^{\mu}_{23}(u_2-u_3)
{\sf R}^{\lambda}_{13}(u_1-u_3)
{\sf R}_{12}^{\lambda \mu}(u_1 -u_2)
\eeq
This Yang-Baxter relation generalizes (\ref{R2}).
Here space 1 is $V^{(\lambda )}$, space 2 is
$V^{(\mu )}$ and space 3 is $\CC ^N$. The explicit form
of $R^{\lambda \mu}(u)$ is much more complicated than
(\ref{R4}). It can be obtained from the
universal $R$-matrix for the quantum affine algebra
$U_q(\widehat{gl}(N))$ \cite{KhorTolst} by specifying it to
finite-dimensional evaluation
representations or by the fusion procedure
\cite{KRS81,fusion1,ACDFR06} applied to the
fundamental $R$-matrix ${\sf R}(u)$.
The $g$-invariance (\ref{g1}) is extended to
$R^{\lambda \mu}(u)$ as follows:
\beq\label{g2}
{\sf R}^{\lambda \mu}_{12}(u)\, \pi_{\lambda}({\sf g})_1\,
\pi_{\mu}({\sf g})_2=
\pi_{\lambda}({\sf g})_1 \,
\pi_{\mu}({\sf g})_2 \,{\sf R}^{\lambda \mu}_{12}(u).
\eeq

The higher transfer-matrices, or $T$-operators, are
constructed in a similar way to (\ref{R3}) by taking trace
of the product of $R$-matrices ${\sf R}^{\lambda}(u-u_i)$
in the auxiliary space $V^{(\lambda )}$:
\beq\label{R6}
T^{\lambda}(u)=\mbox{tr}_{V^{(\lambda )}}
\Bigl ( {\sf R}^{\lambda}_{10}(u-u_1){\sf R}^{\lambda}_{20}(u-u_2)\ldots
{\sf R}^{\lambda}_{L0}(u-u_L)\, \pi_{\lambda} ({\sf g})\Bigr ).
\eeq
Here the space with index 0 is the auxiliary space $V^{(\lambda )}$.
These $T$-operators act in the same quantum space
${\cal H}=\CC ^{\otimes L}$.
If $\lambda = (1)$ is the 1-box diagram, then definition
(\ref{R6}) coincides with (\ref{R5}). By analogy,
we will call ${\sf g}$ of the form (\ref{twistel})
the twist element.
The Yang-Baxter equation (\ref{R5}) and the
${\sf g}$-invariance (\ref{g2}) imply that the
$T$-operators with the same ${\sf g}$
commute for all $u$ and $\lambda$:
$[T^{\lambda} (u), \, T^{\mu}(v)]=0$, and can be diagonalized
simultaneously.

An important property of the $T$-operators defined by
(\ref{R6}) is that they vanish identically if the
first column of $\lambda$ is longer than $N$.

Set
$$
H_a = \sum_{l=1}^{L}e_{aa}^{(l)}\,,
\quad \quad e_{aa}^{(l)}=
\underbrace{\II \otimes \ldots \otimes \II}_{l-1}\otimes \,
e_{aa} \otimes
\underbrace{\II \otimes \ldots \otimes \II}_{L-l}\, ,
$$
then the ${\sf g}$-invariance implies that
$[T^{\lambda}(u), \, H_a]=0$. Therefore, the eigenstates
of the transfer matrices can be classified according to eigenvalues
of the operators $H_a$. Let
$$
{\cal H}=\bigoplus _{M_1, \ldots , M_N}{\cal H}(\{M_a\})
$$
be the decomposition of the quantum space ${\cal H}$ into
the direct sum of eigenspaces for the operators $H_a$
with the eigenvalues $M_a\in \ZZ_+$, $a=1, 2, \ldots , N$, then
eigenstates of $T^{\lambda}(u)$ lie in the
spaces ${\cal H}(\{M_a\})$. Since $\sum_a e_{aa}=\II$ is the
unit matrix, $\sum_a H_a = L \, \II^{\otimes L}$ and thus
\beq\label{MMM}
\sum_{i=1}^{N}M_i =L.
\eeq

For the trivial representation
(corresponding to the empty Young diagram $\emptyset$)
$\pi_{\emptyset}(L^{\pm}_{ab}) =1$ if $a=b$ and $0$ otherwise,
$\pi_{\emptyset}({\sf g})=1$.
Formula (\ref{R4}) yields $R^{\emptyset}(u)=e^{\gamma u}-
e^{-\gamma u}$, where multiplication by the unity
matrix $\II$ is
implied. Therefore,
we can define the $T$-operator
for the trivial representation as follows:
\begin{equation}\label{phia}
T^{\emptyset }(u) =2^L\prod_{i=1}^{L}\sinh (\gamma (u-u_i)).
\end{equation}
For the one-dimensional representation $\pi_{(1^N)}$
(corresponding to the Young diagram $(1^N)$ with one column
of height $N$)
$\pi_{(1^N)}(q^{{\sf h_a}}) =q$ for diagonal
generators and $0$ otherwise,
$\pi_{\emptyset}({\sf g})=\det g$.
Formula (\ref{R4}) yields $R^{(1^N)}(u)=e^{\gamma (u +1)}-
e^{-\gamma (u+1)}$, where multiplication by the unity
matrix $\II$ is
implied. Therefore,
the $T$-operator
for the representation with the highest weight $(1^N)$
(the quantum determinant of the quantum monodromy matrix) is
given by:
\begin{equation}\label{phib}
T^{(1^N)}(u) =2^L\det g \prod_{i=1}^{L}\sinh (\gamma (u+1-u_i))=
\det g \, T^{\emptyset}(u+1).
\end{equation}

For general $\lambda$ the $T$-operator is the Laurent polynomial
in $e^{\gamma u}$ of the similar form:
\beq\label{R7}
T^{\lambda}(u)=\sum_{k=-L/2}^{L/2}G^{\lambda}_k \, e^{2k\gamma u}.
\eeq
The coefficients $G^{\lambda}_k$ of the $T$-operators
with fixed ${\sf g}$ generate the full family of commuting operators
(the Bethe algebra of the vertex model).

The operators $T^{\lambda}(u)$ appear to be functionally dependent.
They are known to obey some functional relations which are given by
the Cherednik-Bazhanov-Reshetikhin (CBR) determinant formulas
\cite{Chered,BR90}. These formulas express $T^{\lambda}(u)$ for
arbitrary $\lambda$ through the transfer matrices
$T_s (u):= T^{(s)}(u)$ corresponding to 1-row diagrams of length
$s$ or through the transfer matrices
$T^a (u):= T^{(1^a)}(u)$ corresponding to 1-column diagrams of height
$a$:
\begin{equation}\label{trm1}
T^{\lambda}(u)
=\displaystyle{
\Bigl ( \prod_{k=1}^{\lambda_{1}'-1}
T^{\emptyset}(u\! -\! k)\Bigr )^{-1}\!\!
\det_{i,j =1,\ldots , \lambda_1^{\prime}}
T_{\lambda_i  - i + j}(u\! -\! j\! +\! 1)},
\end{equation}
\begin{equation}\label{trm2}
T^{\lambda}(u)
=\displaystyle{
\Bigl ( \prod_{k=1}^{\lambda_{1}-1}
T^{\emptyset}(u\! +\! k)\Bigr )^{-1}\!\!
\det_{i,j =1,\ldots , \lambda_1}
T^{\lambda '_i  - i + j}(u\! +\! j\! -\! 1)}.
\end{equation}
Hereafter $\lambda'$ denotes the transposed diagram (with respect to
the main diagonal), so that $\lambda_1'$
is the height of the first column, and $\emptyset $ is the empty
diagram. One can show that formulas (\ref{trm2}) follow from
(\ref{trm1}) and vice versa.

\section{The master $T$-operator}

Let ${\bf t}=\{t_1, t_2, t_3, \ldots \}$ be an infinite set
of parameters which we call times because they will
have the meaning of hierarchical times in the MKP hierarchy.
The Schur polynomials $s_{\lambda}({\bf t})$
labeled by Young diagrams $\lambda$ can be defined
by the determinant formula
\beq\label{T1}
s_{\lambda}({\bf t})=\det_{i,j=1, \ldots , \lambda_1'}
h_{\lambda_i -i +j}({\bf t}),
\eeq
where the polynomials $h_j$ are defined with the help of
the generating series
$$
\exp \Bigl ( \sum_{k\geq 1}t_k z^k\Bigr )=
1+ h_1({\bf t})z +h_2({\bf t})z^2 +\ldots
$$
It is convenient to put $h_0({\bf t})=1$, $h_{n}({\bf t})=0$ for $n<0$
and $s_{\emptyset}({\bf t})=1$.
The functions $h_j$ are elementary Schur polynomials in the sense that
for 1-row diagrams $\lambda =(j)$ with $j$ boxes
$s_{(j)}({\bf t})=h_j ({\bf t})$. Equivalently, one can define
\beq\label{T2}
s_{\lambda}({\bf t})=\det_{i,j=1, \ldots , \lambda_1}
e_{\lambda'_i -i +j}({\bf t}),
\eeq
where the polynomials $e_j$ are defined with the help of
the generating series
$$
\exp \Bigl (\sum_{k\geq 1}(-1)^{k-1}t_k z^k\Bigr )=
1+ e_1({\bf t})z +e_2({\bf t})z^2 +\ldots
$$
For 1-column diagrams $\lambda =(1^j)$ with $j$ boxes
$s_{(1^j)}({\bf t})=e_j ({\bf t})$. Equations (\ref{T1}),
(\ref{T2}) are known as Jacobi-Trudi formulas.
It can be proved \cite{Macdonald} that the Schur
polynomials form a basis in the space of symmetric functions
of the variables $x_i$ defined by $kt_k =\sum_i x_i^k$.

We note the Cauchy-Littlewood
identity
\beq\label{id}
\sum_{\lambda}s_{\lambda}({\bf t})s_{\lambda}({\bf t}')=
\exp \Bigl ( \sum_{k\geq 1}kt_k t'_k\Bigr ),
\eeq
where the sum is over all Young diagrams including the empty one.
Writing it in the form
$$
\sum_{\lambda}s_{\lambda}({\bf y})s_{\lambda}(\tilde \p )=
\exp \Bigl ( \sum_{k\geq 1}y_k \p_{t_k}\Bigr ),
$$
where $\tilde \p =\{\p_{t_1} , \frac{1}{2}\p_{t_2}, \frac{1}{3}\p_{t_3},
\ldots \, \}$
and applying to $s_{\mu}({\bf t})$, we get:
\beq\label{id1}
\left. \phantom{\int}
s_{\lambda}(\tilde \p )s_{\mu}({\bf t})\right |_{{\bf t}=0}=
\delta_{\lambda \mu}\,.
\eeq

Following \cite{AKLTZ11},
we introduce a generating function of the $T$-operators
(the master $T$-operator) depending on the infinite number of
parameters ${\bf t}=\{t_1, t_2, \ldots \}$:
\begin{equation}\label{master}
T(u,{\bf t})=\sum_{\lambda}s_{\lambda}({\bf t})T^{\lambda }(u).
\end{equation}
These operators commute for different values of the
parameters: $[T(u,{\bf t}),\, T(u',{\bf t'})]=0$.
Since $T^{\lambda }(u)=0$ if $\lambda_1'>N$, the sum in
(\ref{master}) is actually restricted to diagrams with
$\lambda_1'\leq N$.
The $T$-operators $T^{\lambda}(u)$ can be restored from the
master $T$-operator according to the formula
\begin{equation}\label{master33}
\left. \phantom{\int}
T^{\lambda}(u)=s_{\lambda}(\tilde \p )T(u,{\bf t})\right |_{{\bf t}=0},
\end{equation}
which follows from (\ref{id1}). In particular,
\beq\label{empty}
T^{\emptyset}(u)=T(u,0),
\eeq
\beq\label{T(u)}
\left. \phantom{\int}T(u)=T^{(1)}(u)=
\p_{t_1}T(u,{\bf t})\right |_{{\bf t}=0}.
\eeq

Below we use the standard notation
\beq\label{stand1}
{\bf t}\pm [z^{-1}]= \Bigl \{ t_1 \pm z^{-1}, \,
t_2 \pm \frac{1}{2}\, z^{-2}, \,
t_3 \pm \frac{1}{3}\, z^{-3}, \, \ldots \Bigr \},
\eeq
\beq\label{stand2}
\xi ({\bf t}, z)=\sum_{k=0}^{\infty}t_k z^k\,.
\eeq
Eq. (\ref{master33}) implies
that $T(u, 0\pm [z^{-1}])$ is the
generating series for $T$-operators corresponding to the
1-row  and 1-column diagrams respectively:
\beq\label{master44}
T(u, [z^{-1}])=\sum_{s\geq 0}z^{-s}T_s(u), \quad
T(u, -[z^{-1}])=\sum_{a\geq 0}(-z)^{-a}T^a(u).
\eeq

As it was proven in \cite{AKLTZ11}, the CBR formulas (\ref{trm1})
imply that the master $T$-operator obeys the bilinear identity
\begin{equation}\label{bi2a101}
\oint_{{\cal C}} e^{\xi ({\bf t}-{\bf t'}, z)}
z^{u-u'} T(u, {\bf t} -[z^{-1}])\, T(u', {\bf t'} +[z^{-1}]) \, dz=0
\end{equation}
for all $u,u', {\bf t}, {\bf t'}$. The contour ${\cal C}$
encircles the cut between $0$ and $\infty$.
By standard manipulations \cite{DJKM83,JM83},
one can derive from (\ref{bi2a101}) the
infinite KP and MKP hierarchies of differential (in $t_k$'s)
and differential-difference (in $t_k$'s and $u$) equations.
The variable $u$ is the so-called ``zero time''; it is
naturally included in the extended sequence of times
$t_0=u, t_1, t_2 , \ldots$.
Choosing $u,u', {\bf t}, {\bf t'}$ in a special way, one can also
derive from (\ref{bi2a101}) the following bilinear equations:
\begin{equation}\label{bi2a}
\begin{array}{c}
(z_2-z_3)T\left (u, {\bf t}+[z_{1}^{-1}]\right )
T\left (u,{\bf t} +[z_{2}^{-1}]+[z_{3}^{-1}]\right )
\\ \\
\hspace{1cm}+ \,\, (z_3-z_1)T\left (u, {\bf t}+[z_{2}^{-1}]\right )
T\left (u,{\bf t} +[z_{1}^{-1}]+[z_{3}^{-1}]\right )
\\ \\
\hspace{2cm}+  \,\, (z_1-z_2)T\left (u, {\bf t}+[z_{3}^{-1}]\right )
T\left (u,{\bf t} +[z_{1}^{-1}]+[z_{2}^{-1}]\right ) =0,
\end{array}
\end{equation}

\begin{equation}\label{bi3a}
\begin{array}{c}
z_2 T\left (u+1, {\bf t}+[z_{1}^{-1}]\right )
T\left (u, {\bf t}+[z_{2}^{-1}]\right )
-z_1 T\left (u+1, {\bf t}+[z_{2}^{-1}]\right )
T\left (u, {\bf t}+[z_{1}^{-1}]\right )
\\ \\
+\,\, (z_1 -z_2)T\left (u+1, {\bf t}+[z_{1}^{-1}] +[z_{2}^{-1}]\right )
T\Bigl (u,{\bf t}\Bigr )=0.
\end{array}
\end{equation}
They are known as Hirota or Hirota-Miwa equations for the
$\tau$-function \cite{Hirota81,Miwa82}. In this sense the
master $T$-operator (any of its eigenvalues) is the $\tau$-function
of the classical MKP hierarchy (see, e.g., \cite{TakTeo06}).
Equation (\ref{master}) can be regarded as the Schur function
expansion of the $\tau$-function (see also \cite{Schur}). 

Note that the transformation
$$T(u, {\bf t})\rightarrow
C(u)\exp \Bigl (\sum_k c_k t_k \Bigr )T(u, {\bf t})$$
with arbitrary function $C(u)$ and arbitrary
constant coefficients $c_k$ preserves the space of
$\tau$-functions. Two $\tau$-functions are regarded as
essentially different if they are not obtained from each
other by such transformation.

As a function of $u$, the master $T$-operator has the
structure similar to (\ref{R7}):
\beq\label{R7a}
T(u, {\bf t})=\sum_{k=-L/2}^{L/2}G_k({\bf t}) \, e^{2k\gamma u}.
\eeq
In particular, the highest and the lowest coefficients
$G_{\pm L/2}({\bf t})$ are easy to calculate. For example,
the highest coefficient $G_{L/2}^{\lambda}({\bf t})$ is
$$
G_{L/2}^{\lambda}({\bf t})=
\exp \Bigl ( -\gamma \sum_{n=1}^{L}u_n\Bigr )
\sum_{a_i, b_i}\mbox{tr}_{V^{(\lambda )}}\left (
L_{a_1 b_1}^{+}L_{a_2 b_2}^{+}\ldots L_{a_L b_L}^{+} \right )
e_{a_1b_1}^{(1)}e_{a_2b_2}^{(2)}\ldots e_{a_Lb_L}^{(L)}
$$
(one should take the first terms from each $R$-matrix
(\ref{R4})). Since all matrices $L_{ab}^{+}$ here
are upper triangular, the trace is equal to that of the product of
diagonal matrices with the same diagonal elements:
$$
\begin{array}{lll}
G_{L/2}^{\lambda}({\bf t})&=&
\displaystyle{
\exp \Bigl ( -\gamma \sum_{n=1}^{L}u_n\Bigr )
\sum_{a_i}\mbox{tr}_{V^{(\lambda )}}\left (
q^{{\sf h}_{a_1}+\ldots +
{\sf h}_{a_L}}p_1^{{\sf h}_{1}}\ldots p_L^{{\sf h}_{L}}\right )
e_{a_1b_1}^{(1)}\ldots e_{a_Lb_L}^{(L)}}
\\ &&\\
&=&\displaystyle{\!\!\!
\sum_{M_1, \ldots , M_L}\!\!
\exp \Bigl ( -\gamma \!\!\sum_{n=1}^{L}u_n\Bigr )
\mbox{tr}_{V^{(\lambda )}}\!\!
\left ((e^{\gamma M_1}p_1)^{{\sf h}_1}\ldots
(e^{\gamma M_L}p_L)^{{\sf h}_L}\right )
\!\! \sum_{a_i:\{M_j\}}
e_{a_1a_1}^{(1)}\ldots e_{a_La_L}^{(L)}},
\end{array}
$$
where the last sum goes over all sequences of indices
$a_1, \ldots a_L$ such that the number of indices equal to
$j$ is $M_j$. It is easy to see that
$$
\sum_{a_i:\{M_j\}}
e_{a_1a_1}^{(1)}\ldots e_{a_La_L}^{(L)}=
\frac{L!}{M_1! \ldots M_L!}\,
P_{M_1 , \ldots , M_N},
$$
where $P_{M_1 , \ldots , M_N}$ is the projector to the subspace
${\cal H}(\{M_i\})$.
Set $\displaystyle{y_k=\frac{1}{k}\sum_{a=1}^{N}
e^{\gamma M_ak}p_a^k}$, then
$$
\mbox{tr}_{V^{(\lambda )}}\left ((e^{\gamma M_1}p_1)^{{\sf h}_1}\ldots
(e^{\gamma M_L}p_L)^{{\sf h}_L}\right )=
s_{\lambda}({\bf y}).
$$
The calculation for $G_{-L/2}({\bf t})$ is similar.
Using the Cauchy-Littlewood
identity (\ref{id}), we get:
\beq\label{R8}
\left. \phantom{\int}G_{\pm L/2}({\bf t})\right |_{{\cal H}(\{M_i\})}
=\frac{(\pm 1)^{L}L!}{M_1! \ldots M_L!}\,
\exp \left (
\mp \gamma \sum_{n=1}^{L}u_n \, +\,
\sum_{k\geq 1}\sum_{a=1}^{N}t_k p_a^k e^{\pm \gamma kM_a}\right ).
\eeq

Let $\left |\omega \right >=
\left |\omega (\{M_i\})\right >\in {\cal H}(\{M_i\})$ be an eigenstate
of $T(u, {\bf t})$,
$$
T(u, {\bf t})\left |\omega (\{M_i\})\right >=
\left |\omega (\{M_i\})\right >
\tau_u ({\bf t}; \left |\omega (\{M_i\})\right >),
$$
then the corresponding eigenvalue
$\tau_u ({\bf t}; \left |\omega (\{M_i\})\right >)$
can be written
in the form
\beq\label{T3}
\tau_u ({\bf t}; \left |\omega (\{M_i\})\right >)=C({\bf t})
\prod_{k=1}^{L}\sinh (\gamma (u-u_k({\bf t})).
\eeq
We will call the expression in the right hand side
a trigonometric polynomial (of degree $L$).
The common multiplier $C({\bf t})$ and the
roots of this trigonometric polynomial depend on
all the times $t_1, t_2 , \ldots$ (and on $\left |\omega \right >$).
Comparing with (\ref{R8}), we find
$$
C({\bf t})=\frac{2^L L!}{M_1! \ldots M_N!}\,
\exp \Bigl ( \sum_{k\geq 1}\sum_{a=1}^{N}
t_k p_a^k \cosh (\gamma M_a k)\Bigr ).
$$
From
(\ref{empty}) and (\ref{phia}) it is clear that
the initial values of these roots are inhomogeneity
parameters at the lattice sites: $u_i(0)=u_i$.

\section{Trigonometric solutions of the MKP hierarchy}

In this section we study solutions of the MKP
hierarchy which are periodic in the variable $t_0=u$
with period $2\pi i/\gamma$. We call them trigonometric solutions.
For solutions of this class, the $\tau$-function is
a ``trigonometric quasi-polynomial" of $u$, i.e., a Laurent
polynomial of the variable $e^{\gamma u}$
possibly multiplied by an exponential function of $u$.

\subsection{The construction of trigonometric solutions}

By trigonometric solutions of the MKP
hierarchy we mean $\tau$-functions
which are
poly\-no\-mi\-als in $e^{\gamma u}$ for some $\gamma$ multiplied by
an exponential function of $u$. They can be viewed
as degenerations of double-periodic (elliptic) solutions in the
complex plane of the variable $u=t_0$ (they correspond to vertex
models with elliptic $R$-matrices). The general theory
of elliptic solutions for the KP hierarchy was developed
in \cite{Krichever-ell} and extended to the MKP hierarchy
in \cite{KZ95}. The trigonometric degeneration simplifies
the construction and makes it more explicit
\cite{Krichever83}.
Here we apply it to the case of our interest.
The $\tau$-function for the trigonometric solutions
will be obtained below in the form of the Casorati determinant
\cite{OHTI93}.

Let $\tau_u({\bf t})$ be the $\tau$-function of the MKP hierarchy.
The Baker-Akhiezer function and its adjoint are defined in the
following way \cite{DJKM83}:
\beq\label{psi1}
\psi_u ({\bf t},z)=z^u e^{\xi ({\bf t}, z)}
\, \frac{\tau_u ({\bf t}-[z^{-1}])}{\tau_u ({\bf t})},
\eeq
\beq\label{psi2}
\psi_u ^*({\bf t},z)=z^{-u} e^{-\xi ({\bf t}, z)}
\, \frac{\tau_u ({\bf t}+[z^{-1}])}{\tau_u ({\bf t})}
\eeq
($\xi ({\bf t}, z)$ is given in (\ref{stand2})). In general, the
ratios of the $\tau$-functions in the right hand sides can be
expanded in infinite series around $z=\infty$:
\beq\label{psi1a}
\psi_u ({\bf t},z)=
z^u e^{\xi ({\bf t}, z)}\left (1+
\frac{w_1({\bf t})}{z}+\frac{w_2({\bf t})}{z^2}+\ldots \right ),
\eeq
\beq\label{psi2a}
\psi_u ^*({\bf t},z)=
z^{-u} e^{-\xi ({\bf t}, z)}\left (
\frac{w_1^*({\bf t})}{z}+\frac{w_2^*({\bf t})}{z^2}+\ldots \right ).
\eeq
According to the Krichever's theory of
general algebro-geometric solutions \cite{Krichever-alggeom},
these solutions can be characterized
and explicitly constructed by fixing certain analytic properties of
the Baker-Akhiezer function on a Riemann surface of the
complex variable $z$ (the classical spectral parameter).
Recall that the quantum spectral parameter $u$ is the zero time
in the classical MKP hierarchy.

For
trigonometric solutions,
the Riemann surface is the Riemann sphere (compactified complex plane),
which represents a genus zero algebraic curve with singularities.
Correspondingly, the Baker-Akhiezer function is, in this case, a rational function on the
complex $z$-plane multiplied by power-like and exponential factors
which give the required
asymptotics (the essential singularity at infinity).
For non-integer $u$ the points $0$ and $\infty$
are branch points for the the Baker-Akhiezer function.
In order to make it single-valued, one should make a cut between
$0$ and $\infty$.

We know that the second series in (\ref{master44})
truncates at $a=N$. This suggests
to assume the following ansatz for the
Baker-Akhiezer function, in which the series in (\ref{psi1a})
truncates at the $N$-th term:
\beq\label{trig1}
\psi_u ({\bf t},z)=z^u e^{\xi ({\bf t},z)}\left (1+\frac{w_1({\bf t})}{z}
+\ldots + \frac{w_N({\bf t})}{z^N}\right ).
\eeq
This explicitly defines the function
$z^{-u} e^{-\xi ({\bf t}, z)}\psi_u ({\bf t},z)$ as a rational function
on the extended complex plane.
The multiplicity $N$ of the pole at $z=0$
is a discrete parameter characterizing the class
of solutions to be constructed.
Fix $N$ points $p_i\in \CC$,
$N$ non-negative integer numbers $M_i$ such that
$$
M_1 +M_2 +\ldots +M_N=L
$$
and the set of parameters
$b_{i,m}$ with $i=1,\ldots , N$, $m=-\frac{1}{2}M_i,
\frac{1}{2}M_i +1,
\ldots , \frac{1}{2}M_i$
(we assume that
$b_{i, \pm \frac{1}{2}M_i}\neq 0$ for all $i$).
Let us impose $N$ conditions of the form
\beq\label{trig2}
\sum_{m=-M_i/2}^{+M_i/2}b_{i,
m}\, \psi_u ({\bf t},p_{i}e^{2\gamma m}) =0\,,
\quad i=1, \ldots , N,
\eeq
which are supposed to hold for any values of $u$, $t_i$.
The sum goes over all integer numbers between $-\frac{1}{2}M_i$ and
$-\frac{1}{2}M_i$ for even $M_i$ and over all half-integer
numbers between $-\frac{1}{2}M_i$ and
$-\frac{1}{2}M_i$ for odd $M_i$.

These conditions yield a system of $N$ linear equations for $N$ coefficients
$w_k$ which allows one to fix the Baker-Akhiezer function $\psi$.
The general theory guaranties that the $\tau$-function associated with
this $\psi$-function according to (\ref{psi1})
solves the MKP hierarchy.
The points $p_i$ and entries of the matrix $b_{i,m}$ are parameters
of the solution.
The coefficients $w_k$ appear to be rational functions
of $e^{\gamma u}$ while the $\tau$-function is a
trigonometric polynomial
in $u$
(possibly
multiplied by an exponential function of $u$).
From the algebro-geometric point of view, these solutions
are associated with singular Riemann surfaces with $N$ ``strings" of
singular points
$$p_ie^{-\gamma M_i}, p_ie^{-\gamma (M_i -2)}, \ldots ,
p_ie^{\gamma (M_i -2)}, p_ie^{\gamma M_i }$$ with the center at $p_i$.
The points of each string are glued with each other in
a complicated way.
Note that the parameters $b_{i,m}$ can be multiplied by
any non-zero complex numbers $k_i$: the transformation
$b_{i,m}\rightarrow k_i b_{i,m}$ does not change anything.

The family of
periodic $N$-soliton solutions is a very particular case of this
construction corresponding to $M_i=1$ for all $i$.
In this case conditions
(\ref{trig2}) become
$b_{i,-1/2}
\psi_u({\bf t}, p_i  e^{-\gamma}
)=-b_{i, 1/2} \psi_u({\bf t}, p_i e^{\gamma})$
and the solutions are associated with the Riemann sphere with
$N$ pairs of double points $p_ie^{\gamma}$ and $p_i e^{-\gamma}$.

It is easy to see that conditions (\ref{trig2}) are equivalent to
the system of linear equations
\beq\label{trig3}
B_{i}(u,{\bf t})+\sum_{k=1}^{N}B_{i}(u-k,{\bf t})w_k=0,
\eeq
where
\beq\label{B}
B_{i}(u,{\bf t}):=p_{i}^{u} \sum_{m=-M_i/2}^{+M_i/2}b_{i,m}({\bf t})
e^{2\gamma m u}\,, \quad
b_{i,m}({\bf t})\equiv b_{i,m} e^{\xi ({\bf t},p_{i}e^{2\gamma m})}.
\eeq
The system can be solved using the
Cramer's rule. This gives the following explicit
expression for the Baker-Akhiezer function:
\beq\label{trig5}
\psi_u ({\bf t},z)=z^u e^{\xi ({\bf t},z )}\,
\frac{\left |\begin{array}{cccc}
1&z^{-1}&\ldots & z^{-N}\\
B_{1}(u,{\bf t})& B_{1}(u\! -\! 1,{\bf t})& \ldots &
B_{1}(u\! -\! N,{\bf t})\\
\vdots & \vdots &\ddots & \vdots \\
B_{N}(u,{\bf t})&B_{N}(u\! -\! 1,{\bf t})& \ldots &B_{N}(u\! -\! N,{\bf t})
\end{array}\right |}{\left |\begin{array}{ccc}
B_{1}(u-1,{\bf t})&\ldots & B_{1}(u-N,{\bf t})\\
\vdots &\ddots & \vdots \\
B_{N}(u-1,{\bf t})& \ldots & B_{N}(u-N,{\bf t})
\end{array}\right |}\,.
\eeq
Comparing with (\ref{psi1}), we conclude, using the obvious property
\beq\label{property}
B_{i}(u,{\bf t}-[z^{-1}])=
B_i(u,{\bf t})-B_i(u+1, {\bf t})z^{-1},
\eeq
that the $\tau$-function is
given by the difference Wronskian (Casorati) determinant
in the denominator:
\beq\label{trig6}
\tau_u ({\bf t})=(\det g)^{-u}
\det_{i,j=1, \ldots , N}B_{i}(u\! -\! j,\, {\bf t})
\eeq
(here $g=\mbox{diag}\, (p_1, \ldots , p_N)$ is the same
matrix as in the previous sections). The factor
$(\det g)^{-u}$ is put here to make $\tau_u ({\bf t})$ a pure
trigonometric polynomial in $u$ of the form (\ref{T3}).
Comparing the highest and the lowest coefficients in (\ref{R7a})
(given by equation (\ref{R8})) with the corresponding coefficients
in (\ref{trig6}), we see that the
parameters $b_{i,\pm \frac{1}{2}M_i}$ obey the following relations:
$$
\Bigl ( \prod_{i=1}^{N}b_{i, \pm \frac{1}{2}M_i}
\Bigr ) (\det g )^{-N} e^{\mp \gamma NL}
\prod_{j<k}\left (p_j e^{\pm \gamma M_j}\! -\!
p_k e^{\pm \gamma M_k}\right )
=\frac{(\pm 1)^{L}L!}{M_1! \ldots M_N!}\, e^{\mp \gamma \sum_{n=1}^{N}
u_n}.
$$

From (\ref{trig5}) it is clear that the last coefficient
in (\ref{trig1}), $w_N$, in terms of the $\tau$-function
is given by
$$
w_N(u, {\bf t})=(-1)^N \, \frac{\tau_{u+1}({\bf t})}{\tau_u({\bf t})}.
$$
We also note the formula
\begin{equation}\label{rat7a}
w_{1}(u,{\bf t})=-\p_{t_1} \log \tau_u({\bf t})
\end{equation}
for the first coefficient
in (\ref{trig1}), $w_1$,
which easily follows from
the obvious relation
\begin{equation}\label{rat4b}
\p_{t_1}B_{i}(u,{\bf t})=B_{i}(u+1,{\bf t}).
\end{equation}

Rewriting (\ref{property}) in the form
$$
B_i(u, {\bf t}+[z^{-1}])=
B_i(u,{\bf t})+z^{-1}B_i(u+1, {\bf t}+[z^{-1}]),
$$
it is straightforward to check that
\begin{equation}\label{rat10}
\tau_u ({\bf t}+[z^{-1}])=(\det g)^{-u}\left |\begin{array}{cccc}
B_{1}\left (u\! -\! 1,{\bf t}\! +\! [z^{-1}]\right )&
B_{1}(u\! -\! 2,{\bf t})&
\ldots & B_{1}(u\! -\! N,{\bf t})\\
B_{2}\left (u\! -\! 1,{\bf t}\! +\! [z^{-1}]\right )&
B_{2}(u\! -\! 2,{\bf t})&
\ldots & B_{2}(u\! -\! N,{\bf t})\\
\vdots & \vdots &\ddots & \vdots \\
B_{N}\left (u\! -\! 1,{\bf t}\! +\! [z^{-1}]\right )&
B_{N}(u\! -\! 2,{\bf t})&
\ldots &B_{N}(u\! -\! N,{\bf t}).
\end{array}\right |
\end{equation}
It directly follows from the definition that
\begin{equation}\label{rat8}
B_{i}(u,{\bf t}+[z^{-1}])=p_i^u
\sum_{m=-M_i/2}^{M_i/2}\frac{b_{i,m}\, z}{z-p_ie^{2\gamma m}}
\, e^{2\gamma m u +
\xi ({\bf t},p_i e^{2\gamma m})}.
\end{equation}
Expanding this in powers of $z^{-1}$, we get:
$$
B_{i}(u,{\bf t}+[z^{-1}]) =B_{i}(u,{\bf t})+
B_{i}(u+1,{\bf t})z^{-1}+B_{i}(u+2,{\bf t})z^{-2}+\ldots
$$
Therefore, the expansion of $\tau_u ({\bf t}+[z^{-1}])$ around
$\infty$ reads
\begin{equation}\label{rat11}
\tau_u ({\bf t}+[z^{-1}])=(\det g)^{-u}\sum_{s=0}^{\infty}z^{-s}
\left |\begin{array}{cccc}
B_{1}\left (u\! +\! s\! -\! 1,{\bf t}\right )&
B_{1}(u\! -\! 2,{\bf t})&
\ldots & B_{1}(u\! -\! N,{\bf t})\\
B_{2}\left (u\! +\! s\! -\! 1,{\bf t}\right )& B_{2}(u\! -\! 2,{\bf t})&
\ldots & A_{2}(u\! -\! N,{\bf t})\\
\vdots & \vdots &\ddots & \vdots \\
B_{N}\left (u\! +\! s\! -\! 1,{\bf t}\right )&B_{N}(u\! -\! 2,{\bf t})&
\ldots &B_{N}(u\! -\! N,{\bf t})
\end{array}\right |.
\end{equation}
We thus see that the adjoint Baker-Akhiezer function has the
determinant representation
\begin{equation}\label{rat12}
\psi^{*}_u({\bf t},z)=z^{-u}e^{-\xi ({\bf t},z)}\,
\frac{\left |\begin{array}{cccc}
B_{1}\left (u\! -\! 1,{\bf t}\! +\! [z^{-1}]\right )&
B_{1}(u\! -\! 2,{\bf t})&
\ldots & B_{1}(u\! -\! N,{\bf t})\\
B_{2}\left (u\! -\! 1,{\bf t}\! +\! [z^{-1}]\right )&
B_{2}(u\! -\! 2,{\bf t})&
\ldots & B_{2}(u\! -\! N,{\bf t})\\
\vdots & \vdots &\ddots & \vdots \\
B_{N}\left (u\! -\! 1,{\bf t}\! +\! [z^{-1}]\right )&
B_{N}(u\! -\! 2,{\bf t})&
\ldots &B_{N}(u\! -\! N,{\bf t})
\end{array}\right |}{\left |\begin{array}{cccc}
B_{1}\left (u\! -\! 1,{\bf t}\right )& B_{1}(u\! -\! 2,{\bf t})&
\ldots & B_{1}(u\! -\! N,{\bf t})\\
B_{2}\left (u\! -\! 1,{\bf t}\right )& B_{2}(u\! -\! 2,{\bf t})&
\ldots & B_{2}(u\! -\! N,{\bf t})\\
\vdots & \vdots &\ddots & \vdots \\
B_{N}\left (u\! -\! 1,{\bf t}\right )&B_{N}(u\! -\! 2,{\bf t})&
\ldots &B_{N}(u\! -\! N,{\bf t})
\end{array}\right |}\, .
\end{equation}
Let us introduce the notation
\begin{equation}\label{rat13a}
\bar B_{k}(u,{\bf t}):= \det_{
{i=1, \,\ldots , \not k , \ldots , N\atop j=1,\, \ldots \, , \, N-1}}
\, B_{i}(u+1-j, \, {\bf t})
\end{equation}
for the minor $M_{k,N}$ of the
$N\! \times \! N$ matrix $B_{i}(u+1-j)$, $1\leq i,j\leq N$.
Then, expanding the determinant in the numerator of
(\ref{rat12}) in the first column, we obtain:
$$
\psi^{*}_u({\bf t},z)=\frac{z^{-u}e^{-\xi ({\bf t},z)}}{\tau_u ({\bf t})}
\sum_{k=1}^{N} (-1)^{k-1}
\bar B_{k}(u-2, {\bf t}) \,B_k (u-1, {\bf t}+[z^{-1}]),
$$
or, substituting (\ref{rat8}),
\beq\label{trig10}
\psi^{*}_u({\bf t},z)=
\frac{z^{-u+1}e^{-\xi ({\bf t},z)}}{(\det g)^u\tau_u ({\bf t})}
\sum_{i=1}^{N}(-1)^{i-1}p_i^{u-1}\!\! \sum_{m=-M_i/2}^{M_i/2}
\frac{b_{i,m}e^{2\gamma m(u-1)+
\xi ({\bf t}, p_ie^{2\gamma m})}}{z-p_i e^{2\gamma m}}\,
\bar B_i(u-2, {\bf t}).
\eeq
This gives the pole expansion of the adjoint Baker-Akhiezer
function. We see that in general it has simple poles at all the
points forming the ``strings''. Below we need this formula rewritten
for the function $\tau_{u}({\bf t}+[z^{-1}])$:
\beq\label{trig11}
\tau_{u}({\bf t}+[z^{-1}])=
z(\det g)^{-u}
\sum_{i=1}^{N}(-1)^{i-1}p_i^{u-1}\!\! \sum_{m=-M_i/2}^{M_i/2}
\frac{b_{i,m}e^{2\gamma m(u-1)+
\xi ({\bf t}, p_ie^{2\gamma m})}}{z-p_i e^{2\gamma m}}\,
\bar B_i(u-2, {\bf t}),
\eeq
with simple poles at the same points.

\subsection{Undressing B\"acklund transformations
for the tri\-go\-no\-met\-ric solutions}

As it was demonstrated in \cite{AKLTZ11}
for models with rational $R$-matrices, the main relations of the
Bethe ansatz method
are naturally built in
the construction of rational solutions to the MKP hierarchy.
The nested Bethe ansatz scheme
appears to be equivalent to a chain of some
special B\"acklund transformations of the initial
rational MKP solution
that ``undress'' it to the trivial solution by reducing the
number of singular points in succession.
All this remains valid for vertex models with trigonometric
$R$-matrices, with the only difference that the undressing
procedure should be applied to the trigonometric solutions.
Technically it becomes even simpler because poles of the adjoint
Baker-Akhiezer function are simple in this case.
In particular, the functions
$\bar B_{k}(u,{\bf t}=0)$ and
$B_{k}(u,{\bf t}=0)$
should be identified, up to some irrelevant factors, with
the (eigenvalues of) the Baxter $Q$-operators on
the first and the last levels of nesting in the nested
Bethe ansatz scheme.

Adding or removing a ``string'' $p_i e^{2\gamma m}$ with the
center at $p_i$
to or from the data
of a trigonometric solution is a B\"acklund transformation.
It sends a trigonometric $\tau$-function to another one.
We will be interested in the removing of a string that results in
decreasing the degree of the trigonometric polynomial
(the undressing transformations).
Basically, such a transformation
can be done by extracting the singular part (the residue)
of the function $\tau_{u}({\bf t}+[z^{-1}])$ at any of its simple
poles which are located at the points of the string with the
center at $p_i$.
Specifically, consider the function
\begin{equation}\label{Back1}
\tau_{u}^{[i]}({\bf t})=(-1)^{i-1}p_{i}^{-1} \!\det g
e^{\gamma M_i (u+1)-
\xi ({\bf t}, p_i e^{-\gamma M_i})}\,
\mbox{res}_{z=p_i e^{-\gamma M_i}}
\tau_{u+1}({\bf t}+[z^{-1}]).
\end{equation}
Equation (\ref{trig11}) implies that
$$
\tau_{u}^{[i]}({\bf t})=b_{i, -\frac{1}{2}M_i}(p_{i}^{-1}\det g)^{-u}
\bar B_i (u-1, {\bf t}),
$$
i.e., up to
the irrelevant constant factor,
it has exactly the same determinant form as $\tau_u ({\bf t})$
with the string with the center at $p_i$ removed.
Therefore, it is
a $\tau$-function, i.e., it
satisfies the same Hirota equations as $\tau_u({\bf t})$ does and
$\tau \rightarrow \tau^{[i]}$ is indeed a B\"acklund
transformation. The degree of the trigonometric polynomial
$\tau_{u}^{[i]}({\bf t})$ is $L-M_i$.
Note that the residue in (\ref{Back1})
is taken at the left edge of the string.
This has an advantage that the coefficient
$b_{i, -\frac{1}{2}M_i}$ is non-zero by definition and thus
the result of the transformation never vanishes identically
(the same holds for the right edge).

The procedure can be continued until one obtains a polynomial
of degree $0$. The inductive definition is as follows.
Fix a set $I_n=\{ i_1, \ldots , i_n\}\subset \{1,2, \ldots , N\}$.
Suppose we have a $\tau$-function
$\tau_{u}^{[i_1 i_2 \ldots i_{n-1}]}({\bf t})$
obtained at the $(n-1)$-th step,
then the $\tau$-function at the $n$-th step is defined as
\beq\label{Back2}
\begin{array}{lll}
\tau_{u}^{[i_1 i_2 \ldots i_{n}]}({\bf t})&=&\displaystyle{
(-1)^{i_n -1}\Bigl (\prod_{j\in \{1, \ldots , N\}\setminus
I_n}\!\!\! p_j\Bigr )
(b_{i_n, -\frac{1}{2}M_{i_n}}\! )^{-1}
e^{\gamma M_{i_n} (u+1)-
\xi ({\bf t}, p_{i_n} e^{-\gamma M_{i_n}})}}
\\ &&\\
&&\,\,\, \times \, \mbox{res}_{z=p_{i_n} e^{-\gamma M_{i_n}}}
\tau^{[i_1 \ldots i_{n-1}]}_{u+1}({\bf t}+[z^{-1}]).
\end{array}
\eeq
This function has the determinant representation
\begin{equation}\label{Back3}
\tau_{u}^{[i_1 i_2 \ldots i_{n}]}({\bf t})=
\Bigl (\prod_{i\in I_n}b_{i, -\frac{1}{2}M_i}\Bigr )
\Bigl (\prod_{j\in \{1,\ldots , N\}\setminus I_n}
p^{-u}_j \Bigr )
\det_{{i=\{1,\ldots , N\}\setminus I_n\atop j=1,\, \ldots \, , \, N-n}}
\, B_{i}(u-j, \, {\bf t}).
\end{equation}

As it is shown in detail in 
\cite{AKLTZ11}, the ``undressing'' chain of B\"acklund transformations
$$
\tau_u({\bf t})\rightarrow
\tau^{[i_1]}_u({\bf t})\rightarrow
\tau^{[i_1 i_2]}_u({\bf t})\rightarrow \ldots \rightarrow
\tau^{[i_1\ldots i_N]}_u({\bf t})\rightarrow 0
$$
is equivalent to the nested Bethe ansatz scheme, with 
the $\tau$-functions $\tau^{[i_1\ldots i_n]}_u({\bf t})$ 
being eigenvalues of 
the master $T$-operators on higher levels of the 
nesting procedure. In particular, $\tau^{[i_1\ldots i_n]}_u({\bf t})$
at ${\bf t}=0$ are
eigenvalues of the Baxter's $Q$-operators. 
They are trigonometric polynomials in $u$ of decreasing 
degree as $n$ increases. This implies the system of Bethe 
equations for their zeros.

\section{Zeros of the master $T$-operator
as the Ruijsenaars-Schneider particles}

As we have seen, eigenvalues of the master $T$-operator
are trigonometric polynomials in the spectral parameter $u$
of the form (\ref{T3}).  The roots of each eigenvalue have
their own dynamics in the times $t_i$. This dynamics is
known \cite{KZ95} to be given by
the trigonometric Ruijsenaars-Schneider model \cite{RuijSch}.
The inhomogeneity parameters
$u_i$ are coordinates of the Ruijsenaars-Schneider particles
at $t_i=0$: $u_i=u_i(0)$.

Here we derive, following \cite{KZ95},
the equations of motion for zeros of the
trigonometric $\tau$-function (the master $T$-operator)
with respect to the first time flow $t_1=t$.
Our starting point is the differential-difference equation
for the Baker-Akhiezer function:
\beq\label{RS1}
\p_{t_1}\psi_u ({\bf t}, z)=\psi_{u+1}({\bf t}, z)+
\p_{t_1} \!\log \! \frac{\tau_{u+1}({\bf t})}{\tau_{u}({\bf t})}
\,\, \psi_u ({\bf t}, z).
\eeq
which follows from the definition and from the Hirota equations.

It is clear from (\ref{psi1}) that $\psi_u({\bf t})$ has
simple poles at $u=u_j({\bf t})$.
Let us introduce the function
$$
\Phi (u, \zeta )=
\frac{\sinh (\gamma (u+\zeta ))}{\sinh (\gamma u)
\sinh (\gamma \zeta )}=\coth (\gamma u) +\coth (\gamma \zeta  )
$$
and adopt
the following pole ansatz for the Baker-Akhiezer function:
\beq\label{RS2}
\psi_u ({\bf t}, z)=z^u \sum_{j=1}^{L} s_j ({\bf t}, z, \zeta))
\Phi (u-u_j({\bf t}, \zeta ).
\eeq
Here $\zeta$ plays the role of an auxiliary spectral parameter.
Substituting this ansatz into (\ref{RS1}), one
is able to derive the equations of motion together with their
Lax representation.
Skipping further details of the calculations, we give the results.
The double poles at $u=u_j$ cancel automatically.
Cancelation of simple poles at $u=u_j-1$ yields:
$$
\gamma \dot u_j \sum_k \Phi (u_{jk}-1, \zeta )s_k =zs_j,
\quad j=1, \ldots , L,
$$
where $\dot u_j := \p_{t_1}u_j$, $u_{jk}:=u_j -u_k$.
Cancelation of simple poles at  $u=u_j$ yields:
$$
\dot s_{j}=\gamma \dot u_j\sum_{k\neq j}\Phi (u_{jk}, \zeta )s_k
+\gamma \Bigl [ \coth (\gamma \zeta )\, \dot u_j
+\sum_{k\neq j}\dot u_k (\coth \gamma (u_{jk}) -
\coth (\gamma (u_{jk}\! +\! 1))\Bigr ]s_j.
$$
Finally, comparison of the constant terms at $u\to \pm \infty$
(we assume that $\gamma$ is real positive) yields the condition
$$
\sum_j \dot s_j = z\sum_j s_j
$$
which does not
add any new constraint
because in fact follows from the previously obtained relations.
The conditions obtained above can be written in the matrix form as
\beq\label{RS3}
{\cal L}(\zeta )\, {\bf s}=z {\bf s}, \quad \quad
\dot {\bf  s} ={\cal M}(\zeta )\, {\bf s},
\eeq
where ${\bf s}=(s_1 , s_2 , \ldots , s_L)^{\rm t}$ and the
matrices ${\cal L}(\zeta )$,  ${\cal M}(\zeta )$ are defined as
\beq\label{RS4}
{\cal L}_{jk}(\zeta )=\gamma \dot u_j \Phi (u_{jk}-1, \zeta ),
\eeq
\beq\label{RS5}
{\cal M}_{jk}(\zeta )=\gamma \Bigl [
(\coth (\gamma \zeta )\! -\! \coth \gamma )\dot u_j +
\sum_{l\neq j}V(u_{jl})\Bigr ] \delta_{jk}+\gamma (1-\delta_{jk})
\dot u_j \Phi (u_{jk}, \zeta ),
\eeq
where
$$
V(u):= \coth (\gamma u)-\coth (\gamma (u+1)).
$$
The compatibility of equations (\ref{RS3}) implies the
Lax equation
\beq\label{RS6}
\dot {\cal L}(\zeta )=[{\cal M}(\zeta ), \, {\cal L}(\zeta )].
\eeq
A direct calculation shows that it
is equivalent to the equations of motion
for the Ruij\-se\-naars-\-Schnei\-der system:
\beq\label{RS7}
\begin{array}{lll}
\ddot u_j&=&\displaystyle{\gamma \dot u_j \sum_{k\neq j}
\dot u_k \Bigl ( V(u_{jk})\! -\! V(u_{kj})\Bigr )}
\\ &&\\
&=&\displaystyle{
-  2\gamma \sinh ^2 \! \gamma  \sum_{k\neq j}
\frac{ \dot u_j \dot u_k
\cosh (\gamma u_{jk})}{\sinh (\gamma (u_{jk}\! -\! 1))
\sinh (\gamma u_{jk})\sinh (\gamma (u_{jk}\! +\! 1))}}\,.
\end{array}
\eeq
In the course of the calculation, the following identities
are useful:
$$
\Phi (u-1, \zeta )\Phi (v, \zeta )-\Phi (u, \zeta )
\Phi (v-1, \zeta )=\Phi (u+v-1, \zeta )\Bigl (
V(-u)-V(-v)\Bigr ),
$$
$$
\p_u \Phi (u-1, \zeta )=\gamma \Bigl (
\coth (\gamma \zeta )-\coth \gamma -V(-u)\Bigr ) \Phi (u-1, \zeta )-
\gamma \Phi (-1, \zeta )\Phi (u, \zeta ).
$$
We also note that the system (\ref{RS6}) is a Hamiltonian
system with the Hamiltonian
\beq\label{RS8}
{\cal H}_1=\sum_{j=1}^{L}e^{v_j}
\prod_{k\neq j}
\left ( \frac{\sinh (\gamma (u_{jk}+1))
\sinh (\gamma (u_{jk}-1))}{\sinh ^2 \! (\gamma u_{jk}))}\right )^{1/2}
\eeq
and the canonically conjugate variables $v_j, u_j$ with the
Poisson brackets $\{v_j, u_k\}=\delta_{jk}$. There are 
also higher Hamiltonians ${\cal H}_j$ in involution which
generate the higher flows with respect to $t_j$.

The spectral curve is given by the equation
\beq\label{curve}
\det_{L\times L}\Bigl ({\cal L}(\zeta )-z\Bigr )=0.
\eeq
One can show that this curve
is the Riemann sphere with points of each string
$p_ie^{2\gamma m}$ being glued in a complicated way.
The coefficients of the characteristic polynomial in the 
l.h.s. are integrals of motion for the Ruijsenaars-Schneider 
system.

Finally, let us stress the specific way of posing the problem 
in the context of the Ruijsenaars-Schneider 
system that corresponds to solution of the vertex model or 
quantum spin chain.
The standard mechanical problem is: given initial coordinates 
and velocities of the particles $u_j(0)$, $\dot u_j(0)$, find
the time evolution $u_j(t)$. By contrast, in order to find 
eigenvalues of the transfer matrix, one should pose the problem
in the following non-standard way: given initial coordinates
$u_j=u_j(0)$ and values of all higher integrals of motion ${\cal H}_j$,
find initial velocities $\dot u_j(0)$. Indeed, the initial velocities
allow one to restore the transfer matrix via residues at its poles:
\beq\label{RS9}
\left. 
\mbox{res}_{u=u_k}\, 
\frac{T(u)}{T^{\emptyset}(u,0)}\right |_{\left |
\omega (\{M_i\})\right >}
=-\dot u_k(0).
\eeq
The solution is not unique: different possible solutions to this
problem correspond to different eigenstates of the transfer matrix
in the sector ${\cal H}(\{M_i\})$.

\section*{Acknowledgments}

The author thanks A.Alexandrov, A.Gorsky, V.Kazakov,
S.Khoroshkin,
I.Krichever, S.Leurent, A.Orlov, T.Ta\-ke\-be and Z.Tsuboi
for discussions. 
Some of these results were reported at the 
workshop ``Classical and
Quantum Integrable Systems'' (Dubna, January
2012).
This work was supported in part by RFBR grant
11-02-01220, by joint RFBR grants 12-02-91052-CNRS,
12-02-92108-JSPS, by grant NSh-3349.2012.2 for support of 
leading scientific schools and
by Federal Agency for Science and Innovations of Russian Federation
under contract 14.740.11.0081.


\begin{thebibliography}{99}



\bibitem{AKLTZ11} A. Alexandrov, V. Kazakov, S. Leurent,
Z. Tsuboi, A. Zabrodin, {\it Classical tau-function for quantum
spin chains}, arXiv:1112.3310.



\bibitem{KLWZ97}
I.\ Krichever, O.\ Lipan, P.\ Wiegmann and A.\ Zabrodin,
{\it Quantum Integrable Models and Discrete Classical Hirota Equations},
Commun. Math. Phys. {\bf 188} (1997) 267-304 [arXiv:hep-th/9604080].

\bibitem{Z97}
A. Zabrodin, {\it
Discrete Hirota's equation in quantum integrable models},
Int. J. Mod. Phys. {\bf B11} (1997) 3125-3158;\\
A. Zabrodin, {\it Hirota equation and Bethe ansatz}, 
Teor. Mat. Fyz., {\bf 116} (1998) 54-100 
(English translation:
 Theor. Math. Phys. {\bf 116}
(1998) 782-819).

\bibitem{KSZ08}
V.~Kazakov, A.~S.~Sorin and A.~Zabrodin,
{\it Supersymmetric Bethe ansatz and Baxter equations from discrete Hirota dynamics},
Nucl.\ Phys.\ B {\bf 790} (2008) 345-413
[arXiv:hep-th/0703147];\\
A. Zabrodin, {\it B\"acklund transformations for
difference Hirota equation and
supersymmetric Bethe ansatz},
Teor. Mat. Fyz. {\bf 155} (2008) 74-93 (English
translation: Theor. Math. Phys. {\bf 155} (2008) 567-584)
[arXiv:0705.4006].

\bibitem{KLT10} V. Kazakov, S. Leurent and Z. Tsuboi,
{\it Baxter's $Q$-operators and operatorial B\"acklund flow
for quantum (super)-spin chains}, Commun. Math. Phys.
{\bf 311} (2012) 787-814 [arXiv:1010.4022].

\bibitem{RuijSch} S. Ruijsenaars and H. Schneider, {\it A new class of
integrable systems and its relation to solitons}, Ann. Phys.
{\bf 170} (1986) 370-405.


\bibitem{MTV} E. Mukhin, V. Tarasov and A. Varchenko,
{\it Gaudin Hamiltonians generate the Bethe algebra
of a tensor power of vector representation of $gl_N$},
St. Petersburg Math. J. {\bf 22} (2011) 463-472
[arXiv:0904.2131];
\\
E. Mukhin, V. Tarasov and A. Varchenko,
{\it KZ characteristic variety as the zero set of
classical Calogero-Moser Hamiltonians}
[arXiv:1201.3990].

\bibitem{Chered} I. Cherednik, {\it An analogue of character formula
for Hecke algebras}, Funct. Anal. and Appl. {\bf 21:2} (1987) 94-95
(translation: pgs 172-174).

\bibitem{BR90} V. Bazhanov and N. Reshetikhin,
{\it Restricted solid-on-solid models connected with simply laced
algebras and conformal field theory}, J. Phys. A: Math. Gen.
{\bf 23} (1990) 1477-1492.

\bibitem{CP}
V. Chari and A. Pressley, {\it A guide to quantum groups},
Cambridge University Press, 1994.

\bibitem{Rosso}
M. Rosso,
{\it Finite dimensional representations of the quantum analogue
of the enveloping algebra
of a complex simple Lie algebra}, Commun. Math. Phys.
{\bf 117} (1988) 581.

\bibitem{KS} A. Klimyk and K. Schm\"udgen, {\it Quantum
groups and their representations},
Springer-Verlag, Berlin, Heidelberg, 1997.

\bibitem{ACDFR06} D. Arnaudon, N. Crampe, A. Doikou, L. Frappat, E. Ragoucy,
{\it Spectrum and Bethe ansatz equations for the
$U_{q}(gl(N))$ closed and open spin chains in any representation},
Ann. H. Poincare {\bf 7} (2006) 1217
[arXiv:math-ph/0512037]



\bibitem{FRT} L. Faddeev, N. Reshetikhin and L. Takhtajan,
{\it Quantization of Lie groups and Lie
algebras}, Algebra and Analysis, {\bf 1} (1989) 178-206
(translation: Leningrad Math. J. {\bf 1} (1990) 193).



\bibitem{KhorTolst} S. Khoroshkin and V. Tolstoy,
{\it Universal $R$-matrix for quantized (super)algebras},
Commun. Math. Phys. {\bf 141} (1991) 599-617.

\bibitem{KRS81}
P. Kulish, N. Reshetikhin and E. Sklyanin,
{\it Yang-Baxter equation and representation theory},
Lett. Math. Phys. {\bf 5} (1981) 393-403

\bibitem{fusion1} I. Cherednik, {\it Special bases of irreducible
representations of a degenerate affine Hecke algebra},
Funk. Anal. i ego Pril. {\bf 20} (1986) 87-88
(translation: Functional Analysis and Its Applications,
{\bf 20} (1986) 76-78).


\bibitem{Macdonald} I. Macdonald, {\it Symmetric functions and
Hall polynomials}, 2nd ed., Oxford University Press, 1995.

\bibitem{DJKM83} E. Date, M. Jimbo, M. Kashiwara and T. Miwa,
{\it Transformation groups for soliton equations},
in "Nonlinear integrable systems -- classical and quantum",
eds. M. Jimbo and T. Miwa, World Scientific, pp. 39-120 (1983).

\bibitem{JM83} M. Jimbo and T. Miwa, {\it Solitons and
infinite dimensional Lie algebras}, Publ. RIMS, Kyoto Univ.
{\bf 19} (1983) 943-1001.

\bibitem{Hirota81}
R. Hirota, {\it Discrete analogue of a generalized Toda
equation}, J. Phys. Soc. Japan {\bf 50} (1981) 3785-3791.

\bibitem{Miwa82}
T. Miwa, {\it On Hirota's difference equations},
Proc. Japan Acad. {\bf 58} (1982) 9-12.

\bibitem{TakTeo06} T. Takebe and L.-P. Teo, {\it Coupled modified
KP hierarchy and its dispersionless limit}, SIGMA {\bf 2} (2006)
072 [arXiv:nlin/0608039].




\bibitem{Schur} 
A. Orlov and T. Shiota, {\it Schur function
expansion for normal matrix model and associated discrete matrix
models}, Phys. Lett. {\bf A343} (2005) 384-396; \\
V. Enolski and J. Harnad,
{\it Schur function expansions of KP tau functions associated to algebraic curves},
Uspekhi Mat. Nauk {\bf 66:4} (2011) 137-178
(Russian Math. Surveys {\bf 66:4} (2011) 767-807),
arXiv:1012.3152.


\bibitem{Krichever-ell} I. Krichever,
{\it Elliptic solutions of the Kadomtsev-Petviashvili equation and
integrable systems of particles},
Funk. Anal. i ego Pril. {\bf 14:4} (1980) 45-54
(translation: Funct. Anal. Appl.,
{\bf 14:4} (1980) 282-290).

\bibitem{KZ95} I. Krichever and A. Zabrodin, {\it
Spin generalization of the Ruijsenaars-Schneider model, non-abelian 2D
Toda chain and representations of Sklyanin algebra},
Uspekhi Mat. Nauk, {\bf 50:6} (1995) 3-56 (translation:
Russ. Math. Surv., {\bf 50:6} (1995) 1101-1150)
[arXiv:hep-th/9505039].

\bibitem{Krichever83}
I.M. Krichever, {\it Rational solutions of the Zakharov-Shabat equations
and completely integrable systems of $N$ particles on a line},
J. Sov. Math., {\bf 21:3} (1983) 335-345;
\\
B.A. Dubrovin, T.M. Malanyuk, I.M. Krichever, V.G. Makhankov,
{\it Exact solutions of a nonstationary
Schrdinger equation with selfconsistent potential},
Sov. J. Part. Nucl. {\bf 19:3} (1988) 579-621.



\bibitem{OHTI93} Y. Ohta, R. Hirota, S. Tsujimoto and
T. Imai, {\it Casorati and discrete Gram type determinant
representations of solutions to the discrete KP hierarchy},
J. Phys. Soc. Japan {\bf 62} (1993) 1872-1886.

\bibitem{Krichever-alggeom} I. Krichever,
{\it Methods of algebraic geometry in the theory of non-linear equations},
Uspekhi Mat. Nauk, {\bf 32:6} (1977) 183-208 (translation:
Russ. Math. Surv., {\bf 32:6} (1977) 185–213).








\end{thebibliography}
\end{document}